\documentclass[aps,amsmath,amssymb,preprintnumbers,groupedaddress,twocolumn]{revtex4}
\usepackage{graphicx,psfrag,hyperref}
\usepackage{amsmath,bbm,array,amsfonts,graphicx,wrapfig,lscape,float,mathtools,multirow,longtable,amssymb}
\usepackage[dvipsnames]{xcolor}

\newcommand{\be}{\begin{equation}}
\newcommand{\ee}{\end{equation}}
\newcommand{\beq}{\begin{equation}}
\newcommand{\beql}[1]{\begin{equation}\label{#1}}
\newcommand{\eeq}{\end{equation}}
\newcommand{\ba}{\begin{array}}
\newcommand{\ea}{\end{array}}
\newcommand{\bea}{\begin{eqnarray}}
\newcommand{\beal}[1]{\begin{eqnarray}\label{#1}}
\newcommand{\eea}{\end{eqnarray}}
\newcommand{\ben}{\begin{enumerate}}
\newcommand{\een}{\end{enumerate}}
\newcommand{\bean}{\begin{eqnarray*}}
\newcommand{\eean}{\end{eqnarray*}}
\newcommand{\eref}[1]{(\ref{#1})}

\newcommand{\tref}[1]{Table~\ref{#1}}
\newcommand{\nn}{\nonumber}

\newcommand{\fref}[1]{Figure \ref{#1}}
\newcommand{\btab}[1]{\begin{tabular}{#1}}
\newcommand{\etab}{\end{tabular}}

\newcommand{\comment}[1]{}

\newcommand{\qed}{\nobreak \ifvmode \relax \else
      \ifdim\lastskip<1.5em \hskip-\lastskip
      \hskip1.5em plus0em minus0.5em \fi \nobreak
      \vrule height0.75em width0.5em depth0.25em\fi}

\newcommand{\Tr}{\text{Tr}}

\definecolor{darkspringgreen}{rgb}{0.09, 0.45, 0.27}
\definecolor{forestgreen}{rgb}{0.13, 0.55, 0.13}

\begin{document}

\preprint{UNIST-MTH-24-RS-02}
\title{Learning BPS Spectra and the Gap Conjecture}

\author{Sergei Gukov$^{a,b}$}
\author{Rak-Kyeong Seong$^{c,d}$}

\email{gukov@math.caltech.edu, seong@unist.ac.kr}

\affiliation{
\it ${}^{a}$ 
Dublin Institute for Advanced Studies, 10 Burlington Rd, Dublin, Ireland, and\\
\it ${}^{b}$
Richard N. Merkin Center for Pure and Applied Mathematics, California Institute of Technology, Pasadena, CA 91125, USA\\
\it ${}^{c}$ 
Department of Mathematical Sciences, and 
\it ${}^{d}$ 
Department of Physics,\\ 
Ulsan National Institute of Science and Technology,\\
50 UNIST-gil, Ulsan 44919, South Korea
}

\begin{abstract}
We explore statistical properties of BPS $q$-series for $3d$ $\mathcal{N}=2$ strongly coupled supersymmetric theories that correspond to a particular family of 3-manifolds $Y$.
We discover that gaps between exponents in the $q$-series are statistically more significant at the beginning of the $q$-series compared to gaps that appear in higher powers of $q$.
Our observations are obtained by calculating saliencies of $q$-series features used as input data for principal component analysis, which is a standard example of an explainable machine learning technique that allows for a direct calculation and a better analysis of feature saliencies.
\end{abstract} 
\maketitle
\noindent

\section{Introduction}

The study of quantum systems and their spectra of quantum states is not only fundamental, but also reveals the underlying quantum structure of spacetime and matter. 
One of the fundamental principles in studying these quantum systems is the relationship between the spectra and the geometry / topology of the space-time in which these quantum systems are defined. 
This relationship is particularly pronounced in the study of black hole physics and other gravitational systems, where guiding principles such as the distance conjecture \cite{Vafa:2005ui, Ooguri:2006in} and the works on black hole microstate counting \cite{Strominger:1996sh} provide profound insights into the quantum architecture of the universe. 

The distance conjecture, a pivotal tool in understanding the properties of the string landscape, suggests that infinite distances in moduli spaces of effective quantum field theories are associated to infinite towers of light states. 
This property provides an argument for a spectral signature of candidate effective field theories in the string landscape that break down over infinite distances in the moduli space of these theories.
The idea that certain quantum states in the spectrum of a theory become critical and cannot be ignored is in line with the importance of counting black hole microstates in consistent theories of quantum gravity.
The work by Strominger and Vafa \cite{Strominger:1996sh}, which connects the counting of black hole microstates to degeneracies in string theory, provides a microscopic explanation of the Beckenstein-Hawking entropy formula.
By doing so this seminal work not only illustrates how the spectrum of black hole microstates can be used to explain macroscopic phenomena like black hole entropy, 
but also underlines the importance of quantum states and why they become critical.

These ideas have their analogues in non-gravitational quantum systems, such as strongly coupled quantum field theories (QFTs). Their spectra often can be conveniently encoded in a $q$-series of the form
\beal{2d3delgenus}
\sum_n q^{E_n} = q^{\Delta} (c_0 + c_1 q + c_2 q^2 + \ldots)
\eea
where $E_n$ plays the role of energy and $\Delta$ is the ``ground state energy.'' For example, the spectrum of a familiar harmonic oscillator would have the $q$-series $q^{1/2} (1 + q + q^2 + \ldots)$, whereas a free chiral boson in $1+1$ dimensions would correspond to $\frac{1}{\eta (q)}$. The latter can be also understood as the elliptic genus of $2d$ supersymmetric QFTs, and higher-dimensional analogues of the elliptic genus --- supersymmetric indices of various kinds --- similarly encode useful information about the spectra (BPS states) in supersymmetric QFTs of dimension $d>2$.

In this paper we focus on a $3d$ variant of the elliptic genus, introduced in \cite{Gadde:2013wq}, that associates a $q$-series of the form \eqref{2d3delgenus} to a $3d$ $\mathcal{N}=2$ supersymmetric QFT with $2d$ $(0,2)$ boundary condition. A large class of strongly-coupled $3d$ $\mathcal{N}=2$ supersymmetric QFTs --- whose Lagrangian description is not known at present --- comes from compactifying $6d$ $(2,0)$ theories on 3-manifolds $Y$. In particular, the complete description of the BPS spectrum is not known for generic $Y$. Yet, for a particular class of $2d$ boundary conditions labeled by Spin$^c$ structures on $Y$, the $2d$-$3d$ analogue of the elliptic genus can be computed for a generic $Y$. The resulting $q$-series, sometimes called Gukov-Pei-Putrov-Vafa (or, GPPV) invariants \cite{Gukov:2016gkn,Gukov:2017kmk}, are denoted by $\hat{Z}_b(q; Y)$, where $b$ labels the choice of a Spin$^c$ structure on $Y$.

In this work, we study BPS state counting captured by the $q$-series $\hat{Z}_b(q; Y)$
of a particular class of 3-manifolds $Y$, which can be defined in terms of a framed graph known as the plumbing graph $\Gamma$ of $Y$.
Our aim is to study the underlying structure of the BPS invariants and, among other things, to understand how the BPS spectrum depends on the structure of the framing coefficients, for a fixed plumbing graph $\Gamma$.
In our analysis we use explainable machine learning techniques that detect the underlying structure of the $q$-series invariants.
By doing so, we discover a fascinating property of the BPS $q$-series that we conjecture to be valid for any collection of plumbed 3-manifolds $Y$: gaps between consecutive $q$-series exponents at the beginning of the series are statistically more significant than exponent gaps that follow.

This observation is somewhat surprising in view of a recent duality, expected to hold for general $\Gamma$, that relates BPS spectra ($Q$-cohomology) of combined $2d$-$3d$ systems described above and chiral logarithmic CFT's in two dimensions \cite{Cheng:2018vpl,Cheng:2022rqr}. 
Indeed, according to Cardy \cite{Cardy:1986ie}, the coefficients $c_n$ in \eref{2d3delgenus} should grow as
\beal{Cardygrowth}
a_n \sim \exp 2\pi \sqrt{\frac{1}{6} c_{\text{eff}} \, n}
\eea
for some constant $c_{\text{eff}}$, called the effective central charge.
And, furthermore, since the effective central charge is the basic characteristic of a $2d$ CFT, one might expect the large-$n$ behavior of the terms in the $q$-series \eref{2d3delgenus} to play a more dominant role in characterizing the theory. Curiously, our finding is essentially the opposite: it is the lower-$n$ terms in the $q$-series / BPS spectrum that capture most information about the quantum system at hand.

By representing the ``gaps''
between consecutive $q$-series exponents as components of a finite dimensional vector, 
we interpret a collection of $q$-series for different 3-manifolds $Y$ 
as a collection of such vectors. 
With this data representation, we then employ dimensional reduction on these vectors via the principal component analysis (PCA) and take a closer look at the resulting covariance matrix and its eigenvalues and eigenvectors, which shed light on the statistical significance of each gap in the $q$-series exponents. 
This application of a standard explainable machine learning technique allows us to identify 
statistically more significant exponent gaps in BPS $q$-series associated to a collection of 3-manifolds $Y$.
As part of the wider study of quantum systems and their spectra of quantum states, 
our work shows that the BPS spectra of $3d$ $\mathcal{N}=2$ theories corresponding to 3-manifolds $Y$ is associated to exponent gaps that are statistically more significant than others in the BPS $q$-series.

Our work also illustrates effectively how techniques from explainable machine learning can be used in order to uncover hidden structures 
in the spectra of states in quantum systems.
Their correspondence to enumerative invariants of the geometry and topology of the spaces in which these quantum systems are defined, makes the application of explainable machine learning techniques extremely fruitful, as shown by our work. 
We emphasize that our application of machine learning techniques in uncovering hidden structures in the BPS spectra of $3d$ $\mathcal{N}=2$ supersymmetric theories associated to plumbed 3-manifolds $Y$ is one of many applications in the quest for hidden structures in quantum systems \cite{He:2017aed,Krefl:2017yox, Ruehle:2017mzq,Carifio:2017bov,Hashimoto:2018ftp,Jejjala:2019kio,Brodie:2019dfx,Ruehle:2020jrk,Gukov:2020qaj,Seong:2023njx,Choi:2023rqg,Dubey:2023dvu,Ri:2023xcn,Gukov:2024buj}.
\\

\begin{figure}[ht!!]
\begin{center}
\resizebox{0.4\hsize}{!}{
\includegraphics[height=5cm]{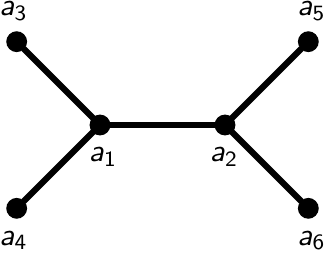} 
}
\caption{
`H-shaped' plumbing graph with framing $\mathbf{m}=(a_1,a_2,a_3,a_4,a_5,a_6)$.
\label{f_fig01}}
 \end{center}
 \end{figure}

\section{Background}

We consider in this work plumbed $3$-manifolds $Y$ characterized by a finite decorated graph with no loops.
The vertices $v$ of the tree are labelled by integers $a_v$, which are the \textit{framing} numbers of the graph.
We call such a finite framed graph associated to $Y$ the \textit{plumbing graph} $\Gamma$.

The $3$-manifold $Y$ given by $\Gamma$ is the boundary of a $4$-manifold $X$, which is also determined by $\Gamma$.
The vertices $V(\Gamma)$ of $\Gamma$ correspond to disk bundles over $S^2$ and the edges $E(\Gamma)$ connecting the 
vertices indicate which disk bundles are glued together to form the $4$-manifold $X$.
Accordingly, the adjacency information of the plumbing tree $\Gamma$ is an essential ingredient for determining the boundary $3$-manifold $Y$ and its associated $4$-manifold $X$.
In addition to the adjacency information, the framing on vertices determines 
how the common boundary on the disk bundles, in this case $\partial D^2 \simeq S^1$, is glued together.
Given that the glueing depends on a relative rotation specified by a winding number, we take the framings on vertices of the plumbing graph connected by an edge as the relative winding number of the corresponding common boundary between two associated glued disk bundles over $S^2$.

\begin{figure}[ht!!]
\begin{center}
\resizebox{0.98\hsize}{!}{
\includegraphics[height=5cm]{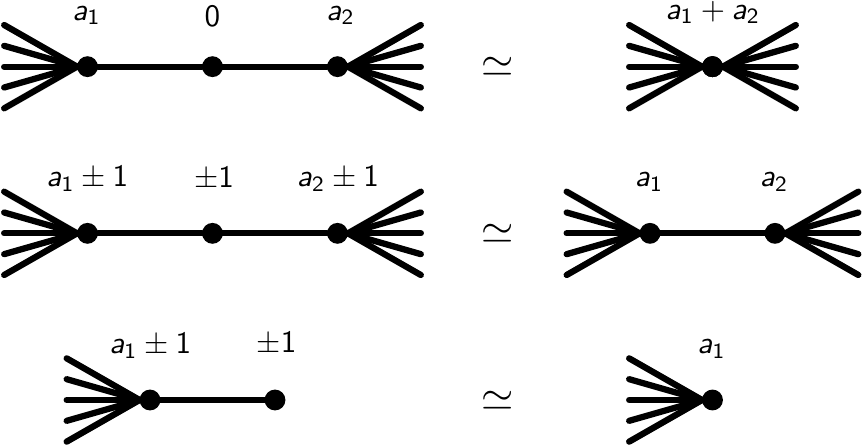} 
}
\caption{
Plumbing graphs related by Neumann moves. 
These plumbing graphs related by these moves result in homeomorphic $3$-manifolds.
\label{f_fig02}}
 \end{center}
 \end{figure}

Following \cite{Gukov:2016gkn,Gukov:2017kmk}, we define the \textit{plumbing matrix} $M_{vw}$ that describes the adjacency of vertices in $\Gamma$ along with framing data,
\beal{es10a01}
M_{vw} = \left\{
\ba{ll}
a_v & \text{~if~ $v=w$} \\
1 & \text{~if~ $(v,w)\in E(\Gamma)$}\\
0 & \text{~otherwise}
\ea
\right.
~,~
\eea
where $E(\Gamma)$ denotes the set of edges in $\Gamma$.
We note that plumbing graphs that are related to each other by \textit{Kirby-Neumann moves} \cite{kirby1978calculus,neumann1981calculus} correspond to the same boundary $3$-manifold $Y$.
These moves are illustrated in \fref{f_fig02}.

By fixing a plumbing tree $\Gamma$ with $|V|$ vertices and an assigned framing on the vertices given by $\mathbf{m} \in \mathbb{Z}^{|V|}$,
we can calculate a generating function of invariants on the $3$-manifold $Y$ known as the \textit{$q$-series} $\hat{Z}_b(q)$ \cite{Gukov:2016gkn,Gukov:2017kmk}.
These invariants are precisely the BPS indices (or, rather, half-indices) associated to the $3d$ $\mathcal{N}=2$ theory obtained from compactifying a $6d$ $(2,0)$ theory on the 3-manifold $Y$.

The $q$-series is given by,
\beal{es10a10}
&&
\hat{Z}_b(q)
=
(-1)^{\pi(M)}
q^{\frac{-3\sigma(M)-\Tr(M)}{4}}
\nn\\
&&
\hspace{0.5cm}
\cdot~
\text{v.p.}
\oint_{|z_v|=1}
\prod_{v\in V(\Gamma)}
\frac{dz_v}{2\pi i z_v}
\left(
z_v - \frac{1}{z_v}
\right)^{2-\deg(v)}
\nn\\
&&
\hspace{4.5cm}
\cdot ~\Theta_b^{-M}(q,\vec{z})
~,~
\eea
where
\beal{es10a11}
\Theta_b^{-M}(q,\vec{z}) = 
\sum_{\vec{l}\in 2M\mathbb{Z}^{|V|} + b}
q^{-\frac{(\vec{l},M^{-1}\vec{l})}{4}}
\prod_{v \in V(\Gamma)} z_v^{l_v} ~.~
\eea
Above, $\text{v.p.}$ refers to the principal value of the integral in \eref{es10a10}, 
which is given by taking the average over the contributions from the integrals over the circles $|z_v| = 1+\epsilon$ and $|z_v| = 1-\epsilon$ for small $\epsilon>0$.
We also have the number of positive eigenvalues for plumbing matrix $M$ denoted by $\pi(M)$ as well as the signature of $M$ given by $\sigma(M)$.

For every plumbed $3$-manifold $Y$, a choice of a Spin$^c$ structure can be described by a characteristic vector $b$,
\beal{es10a15}
b \in (2\mathbb{Z}^{|V|}+ \mathbf{m}) /2M\mathbb{Z}^{|V|} \simeq \text{Spin}^c(Y)~,~
\eea
where $M$ is the plumbing matrix and $\mathbf{m}\in \mathbb{Z}^s$ is the framing vector on $\Gamma$.
In the expression for the $q$-series in \eref{es10a10}, $b$ is precisely the choice of a Spin$^c$ structure on $Y$.

In general, the $q$-series for a $3$-manifold $Y$ characterized by a framed plumbing graph $\Gamma$ takes the form, 
\beal{es10a16}
\hat{Z}_b(q) = q^{\Delta_b} \sum_{n=0}^{\infty} c_n q^n ~,~
\eea
where $\Delta_b \in \mathbb{Q}$ plays the role of the ``ground state energy'' or, in connection with $2d$ logarithmic CFTs (vertex algebras), it can also be thought of as the \textit{conformal weight}.

In this work, we focus on the $q$-series as an invariant of the $3$-manifold $Y$ equipped with a Spin$^c$ structure $b$.
For simplicity, we restrict our study to a subset of $3$-manifolds which have the following additional constraints imposed on them.
(QFTs associated to more general 3-manifolds and the structure of their BPS spectra will appear in a future work.)
\\

\noindent\textit{A Set of $3$-Manifolds $Y$.}
We restrict the plumbing graph to be the graph shown in \fref{f_fig01} with 2 vertices of degree $3$.
The total number of vertices is $|V|=6$.
Given that the number $N_{b}$ of distinct choices for the Spin$^c$ structure is given by 
\beal{es10a20}
N_b = |\det{M}| ~,~
\eea
for simplicity we study integral homology spheres, i.e. 3-manifolds whose plumbing presentation has $\det{M} = 1$, so that there is only one Spin$^c$ structure 
\beal{es10a21}
b= [0,\dots,0] \in \mathbb{Z}^{|V|} ~.~
\eea
We further restrict the set of possible $3$-manifolds $Y$ and their corresponding $q$-series by 
imposing the condition that the eigenvalues of the plumbing matrix $M$ are all negative.

Considering these constraints, we can compute the possible framings $\mathbf{m}$ on the plumbing graph $\Gamma$ in \fref{f_fig01} such that $\Gamma$ remains irreducible under Kirby-Neumann moves. In other words, each such $\Gamma$ can not be reduced to a simpler graph, with only one trivalent vertex.
\tref{t_tab01} summarizes a classification of distinct framings $\mathbf{m}$ on $\Gamma$ up to the maximum value $m_v^{max}$ for $m_v$ for any vertex $v\in V(\Gamma)$. Curiously, before we mod out by the symmetries --- relabeling of the vertices --- the number of possibilities that meet our constraints is huge: $7,529,536$ for $m_v^{max} = 15$ and $47,045,881$ for $m_v^{max} = 20$. However, as \tref{t_tab01} shows, this number is considerably reduced when one mods out by the symmetries.
\\

\begin{table*}[ht!!]
\begin{tabular}{|c|c|c|}
\hline
$m_v^{max}$ & $N_{\mathbf{m}}(m_v^{max})$ & new $\mathbf{m}=(a_1,a_2,a_3,a_4,a_5,a_6)$
\\
\hline\hline
1 & 0 & ---
\\
2 & 0 & ---
\\
3 & 0 & ---
\\
4 & 1 & $(1,3,3,4,3,4)$
\\
5 & 2 & $(1,3,3,5,2,3)$
\\
6 & 2 & ---
\\
7 & 5 & $(1,2,5,7,3,7),(1,4,2,5,2,7),(1,7,2,3,2,3)$
\\
8 & 5 & ---
\\
9 & 5 & ---
\\
10 & 6 & $(1,2,4,9,3,10)$
\\
11 & 8 & $(1,3,2,9,3,11),(1,3,3,4,2,11)$
\\
12 & 8 & ---
\\
13 & 8 & ---
\\
14 & 8 & ---
\\
15 & 8 & ---
\\
16 & 11 & $(1, 2, 3, 8, 11, 16),(1, 2, 3, 11, 5, 16),(1, 2, 5, 16, 2, 7)$
\\
17 & 12 & $(1, 2, 12, 17, 2, 3)$
\\
18 & 13 & $(1, 3, 2, 7, 7, 18)$
\\
19 & 15 & $(1, 2, 4, 17, 2, 19),(1, 3, 2, 11, 2, 19)$
\\
20 & 15 & ---
\\
21 & 16 & $(1, 2, 5, 9, 2, 21)$
\\
22 & 16 & ---
\\
23 & 17 & $(1, 2, 4, 23, 3, 4)$
\\
24 & 17 & ---
\\
25 & 18 & $(1, 2, 4, 25, 2, 11)$
\\
26 & 19 & $(1, 2, 4, 5, 7, 26)$
\\
27 & 20 & $(1, 3, 2, 27, 2, 3)$
\\
28 & 20 & ---
\\
29 & 20 & ---
\\
30 & 20 & ---
\\
\hline
\end{tabular}
\caption{
Framings $\mathbf{m}=[a_1,a_2,a_3,a_4,a_5,a_6]$ for the `H-shaped' plumbing graph with a unique Spin$^c$ structure $b=[0,0,0,0,0,0]$.
The table lists only framings that lead to irreducible framed plumbing graphs $\Gamma$ and a plumbing matrix $M$ with only negative eigenvalues.
In the classification, the framing $m_v$ on a given vertex $v \in V(\Gamma)$ is limited by $m_v \leq m_v^{max}$, and the number of framings satisfying the constraints is given by $N_{\mathbf{m}}(m_v^{max})$ for a bound $m_v^{max}$.
\label{t_tab01}}
 \end{table*}

\begin{table*}[ht!!]
\begin{tabular}{|c|c|}
\hline
$\mathbf{m}$ & $\hat{Z}_b(q)$
\\
\hline\hline
$(1,3,3,4,3,4)$ & 
$q^{1/2} (1+q-q^2-q^4-2 q^5+q^6-q^7-q^8+q^{10}+q^{11}-q^{12}+q^{13}+2 q^{15}+q^{16}-q^{19}+2q^{20}+q^{21}$
\\
& $-q^{23}-q^{25}-q^{27}+q^{28}+q^{29}-2q^{30}-q^{32}-q^{33}-q^{34}+q^{35}-q^{39}-2 q^{40}+q^{42}+q^{44}+\dots)$
\\
&
\\
$(1,3,3,5,2,3)$ & $1/2~q^{1/2} (3+3 q-q^2+2 q^3+q^4-2 q^6-4 q^7+2 q^8-2 q^9-2 q^{10}-3 q^{11}+2 q^{12}+2 q^{13}-6q^{14}-q^{15}$
\\
& $-2 q^{17}-q^{18}+4 q^{21}-2 q^{22}-q^{23}-2 q^{25}+2 q^{26}+4 q^{28}-2q^{29}-2 q^{31}+2 q^{32}+4 q^{33}+2 q^{34}
+ \dots )$
\\
&
\\
$(1,2,5,7,3,7)$ & $1/2~q^{5/2} (1+q^2+q^3+2 q^4+2 q^5-q^6-q^7+2q^9+q^{13}+q^{15}-q^{16}+q^{17}+q^{18}-q^{19}+q^{20}-q^{21}$
\\
&
$-2q^{22}+q^{24}-q^{25}-q^{26}-2q^{27}-q^{29}-q^{30}-q^{31}+q^{32}-q^{35}+q^{36}-q^{38}-2q^{39}-q^{40}+ \dots)$
\\
&
\\
$(1,4,2,5,2,7)$ & $1/2~q^{3/2} (1+2 q-2 q^2+q^3-3 q^4+2 q^5+q^6-3 q^7-q^9+q^{10}-2q^{11}+q^{13}+q^{14}+q^{15}+q^{18}+q^{19}$
\\
& $-q^{20}-q^{22}+q^{25}+q^{28}-q^{30}-q^{31}+q^{32}-q^{37}+q^{41} +q^{42}+q^{44}-2 q^{45}-q^{47}-q^{56}+ \dots)$
\\
&
\\
$(1,7,2,3,2,3)$ & $q^{1/2} (1-2 q+2 q^3+q^4+q^5-2 q^6-q^7-2 q^8-2 q^9+q^{10}+q^{12}+2 q^{13}+3 q^{14}-q^{15}+2q^{16}-2 q^{19}$
\\
& $-2 q^{20}-q^{22}-2 q^{23}-q^{24}+q^{26}+2 q^{27}-2 q^{28}+2 q^{29}+q^{30}+2 q^{31}+2 q^{33}-2 q^{34}-q^{35} +\dots)$
\\
&
\\
$(1,2,4,9,3,10)$ & $1/2~q^{9/2} (1-q+2 q^3-q^6+q^7+q^8-q^{12}+q^{13}-q^{15}-q^{16}+q^{23}-2 q^{31}-2 q^{32}+q^{45}+2q^{49}-q^{51}$
\\
& $+q^{56}-q^{57}+q^{69}+q^{71}+q^{73}-q^{83}-q^{85}-q^{86}-q^{90}-q^{99}-q^{100}+q^{104}+q^{105}-q^{122} + \dots)$
\\
&
\\
$(1,3,2,9,3,11)$ & $1/2~q^{9/2} (1+q+2 q^2-q^3+q^4-q^5-2 q^6+2 q^7-q^8-2 q^9-q^{12}-q^{13}-q^{19}-q^{20}+2 q^{23}+q^{24}$
\\
& $+q^{25}-q^{26}+q^{27}+q^{28}-3 q^{30}+q^{31}-q^{34}+q^{35}-q^{36}-2 q^{37}-q^{38}-q^{39}-q^{40}+2 q^{42}+ \dots)$
\\
&
\\
$(1,3,3,4,2,11)$ & $-1/2~q^{5/2} (1-2 q+q^2-q^6-2 q^7+q^8+4 q^9-q^{10}+q^{12}+q^{13}-2 q^{14}-q^{15}-q^{16}+q^{17}+2 q^{19}$
\\
& $-q^{20}+q^{23}-2 q^{24}-q^{25}-q^{27}-2 q^{29}+q^{31}-q^{33}+3 q^{34}+2q^{35}-q^{36}+q^{40}-q^{41}-q^{42}+ \dots)$
\\
&
\\
$(1,2,3,8,11,16)$ & $1/2~q^{13/2} (2+q^3+q^{10}-q^{12}-2 q^{13}-q^{14}-q^{16}-q^{19}-q^{22}-q^{29}+q^{37}-2 q^{39}+q^{45}+q^{47}$
\\
& $+2q^{52}+q^{54}+q^{57}-q^{60}+q^{64}+q^{65}+q^{67}-q^{68}+q^{69}+q^{71}-q^{72}+q^{75}+q^{79}+q^{81}-q^{83}+ \dots)$
\\
&
\\
$(1,2,3,11,5,16)$ & $1/2~q^{13/2} (1+q^3-q^4+2 q^6-2 q^{11}+q^{20}-q^{21}+q^{23}-2 q^{25}-q^{26}-q^{27}-q^{28}+2 q^{31}-q^{32}$
\\
& $-q^{34}-q^{40}+q^{42}+q^{45}+q^{50}-q^{52}+q^{53}+q^{55}+q^{57}+q^{59}-q^{61}+q^{62}-q^{74}-q^{77}-2 q^{78} +\dots)$
\\
\hline
\end{tabular}
\caption{\textbf{(Part 1)} 
The list of the BPS $q$-series for `H-shaped' plumbing graphs with a unique Spin$^c$ structure $b=[0,0,0,0,0,0]$, and with framing coefficients $\mathbf{m}$ up to $a_v^{max}=29$. 
(For \textbf{Part 2} see \tref{t_tab03}.)
\label{t_tab02}}
 \end{table*}

\begin{table*}[ht!!]
\begin{tabular}{|c|c|}
\hline
$\mathbf{m}$ & $\hat{Z}_b(q)$
\\
\hline\hline
$(1,2,5,16,2,7)$ & $-1/2~q^{11/2} (1-q+q^2+q^3-q^4+2 q^6+2 q^7-q^{12}-q^{13}+q^{15}-2 q^{16}-2 q^{18}-q^{24}-2 q^{25}-2 q^{27}$
\\
& $+2 q^{33}+q^{36}+q^{37}-q^{38}+2 q^{40}-q^{43}-q^{45}+q^{46}-q^{48}-q^{49}-q^{55}+2 q^{56}+q^{61}-2 q^{63}-q^{64}+ \dots)$
\\
&
\\
$(1,2,12,17,2,3)$ & $-1/2~q^{11/2}(2-5 q^7-q^9+q^{10}+q^{15}+q^{20}+2 q^{21}+q^{24}+q^{26}-q^{27}+q^{29}-2 q^{32}+q^{34}+2 q^{35}$
\\
& $+q^{36}-2 q^{37}-3 q^{42}-q^{44}-q^{46}-q^{47}+3 q^{49}-q^{54}-3 q^{56}+q^{59}+q^{61}-q^{65}+q^{66}-2q^{67}+q^{68}+ \dots)$
\\
&
\\
$(1,3,2,7,7,18)$ & $1/2~q^{13/2}(2-2 q^5-q^7-q^9+q^{10}+q^{13}-2 q^{15}+q^{16}+q^{18}-q^{19}-q^{22}+2 q^{25}+q^{36}+q^{38}+q^{40}$
\\
& $-q^{41}+q^{43}-q^{44}-q^{45}-q^{46}-q^{52}-q^{54}+q^{55}+q^{56}-q^{59}+q^{60}+2 q^{63}+q^{65}-q^{67}-q^{69}+ \dots)$
\\
&
\\
$(1,2,4,17,2,19)$ & $-1/2~q^{25/2}(1-q^5+q^6-q^8-q^{11}+q^{12}-2 q^{14}+q^{18}+q^{21}-q^{22}-q^{26}-q^{27}+3 q^{29}-q^{36}+2 q^{39}$
\\
& $-q^{40}+q^{44}+q^{47}+q^{49}+q^{50}-q^{51}+q^{53}+q^{54}+q^{56}-2 q^{57}-q^{60}-q^{62}-q^{63}-2 q^{69}+q^{74}+\dots)$
\\
&
\\
$(1,3,2,11,2,19)$ & $-1/2~q^{11/2}(1-q^2-q^5+2 q^{11}-q^{12}+q^{13}-q^{16}-q^{18}+q^{21}-q^{23}+3 q^{26}-q^{27}+q^{28}-q^{29}+q^{30}$
\\
& $-q^{35}-q^{39}-q^{41}-q^{43}+q^{44}-q^{46}+q^{51}-q^{53}-q^{56}+q^{57}-q^{60}+2 q^{62}+q^{63}-q^{66}+q^{67}+q^{68}+ \dots)$
\\
&
\\
$(1,2,5,9,2,21)$ & $1/2~q^{13/2}(1+q^3-q^4-2 q^5-q^{10}+q^{16}+q^{17}-q^{18}+q^{19}+q^{20}-q^{24}+q^{25}+q^{26}-q^{27}+q^{28}$
\\
& $+q^{29}+q^{32}-q^{35}-q^{36}+q^{37}-q^{39}-2 q^{41}-q^{43}-q^{44}-q^{46}-q^{48}-q^{58}+q^{63}+q^{65}+3 q^{66}+ \dots)$
\\
&
\\
$(1,2,4,23,3,4)$ & $-1/2~q^{11/2}(
1 - 2 q - q^3 - q^6 + q^8 + 2 q^{11} + q^{12} + q^{14} + q^{15} - 
 2 q^{16} + q^{17} - q^{18} + q^{19} - q^{20} + 2 q^{25}
$
\\
& $
-q^{29}+q^{31}-q^{34}-q^{38}-q^{41}-2 q^{42}-q^{44}-q^{48}+q^{49}+q^{61}
+2 q^{66}+q^{67}+q^{69}+q^{71}-q^{74}
+ \dots)$
\\
&
\\
$(1,2,4,25,2,11)$ & $1/2~q^{23/2}(1+2 q^4+q^6-q^7+2 q^{12}-2 q^{13}-q^{16}-q^{18}-2 q^{19}-q^{21}-q^{22}-2 q^{30}+q^{39}-q^{40}+q^{43}$
\\
& $+q^{48}-q^{54}+2 q^{55}-q^{57}+3 q^{61}+2 q^{63}+q^{64}+q^{67}+q^{72}-q^{79}-q^{82}+q^{84}+q^{85}+q^{88}+ \dots)$
\\
&
\\
$(1,2,4,5,7,26)$ & $1/2~q^{25/2}(2+q^6-q^9-4 q^{11}-q^{12}-q^{14}+q^{15}+q^{17}-q^{20}+2 q^{22}+q^{29}+q^{31}-q^{32}+q^{33}-q^{35}$
\\
& $+q^{38}-q^{45}-q^{47}+q^{48}-q^{50}-q^{54}+q^{57}-q^{58}+q^{65}-q^{66}+q^{69}+q^{72}+q^{76}+q^{83}+ \dots)$
\\
&
\\
$(1,3,2,27,2,3)$ & $-1/2~q^{11/2}(1-q-q^2+q^5+q^7-q^{11}+q^{13}-q^{15}-2 q^{16}+q^{17}-q^{21}+q^{22}+2 q^{26}+q^{28}+q^{31}-q^{33}$
\\
& $-q^{35}-q^{37}-q^{38}-q^{40}-q^{42}+q^{46}+q^{49}-q^{50}+q^{54}+2 q^{56}+q^{57}+q^{59}-q^{60}-2 q^{61}-q^{62}+ \dots)$
\\
\hline
\end{tabular}
\caption{\textbf{(Part 2)} 
The list of the BPS $q$-series for `H-shaped' plumbing graphs with a unique Spin$^c$ structure $b=[0,0,0,0,0,0]$, and with framing coefficients $\mathbf{m}$ up to $a_v^{max}=29$. 
(For \textbf{Part 1} see \tref{t_tab02}.)
\label{t_tab03}}
 \end{table*}

\section{Machine Learning BPS $q$-Series}

\noindent
\textit{The $q$-Series and Vector Spaces.}
The $q$-series in the general form shown in \eref{es10a16} can be fully expanded to take the following form,
\beal{es29a01}
\hat{Z}_b(q) = h_0 q^{\Delta_b + s_0} + h_1 q^{\Delta_b + s_1} + h_2 q^{\Delta_b + s_2} + \dots ~,~
\nn\\
\eea
where $h_i \neq 0$ are the non-zero coefficients of the expansions and $s_i \geq 0$ are the integer exponent shifts with $s_0=0$ and $s_i < s_{i+1}$ for all $i$.
When we truncate the expansion up to a maximum order $q^{\Delta_b+s_D}$, the coefficients $h_i$ can be used to define a \textit{vector of invariants} of the following form 
\beal{es29a01b}
\mathbf{h} = [h_0 , h_1, \dots, h_{D}] \in \mathbb{Q}^{D+1} ~.~
\eea
We can also define a \textit{$q$-exponent vector} as
\beal{es29a01b2}
\mathbf{e} = [\Delta_b , \Delta_b + s_1, \dots, \Delta_b + s_D] \in \mathbb{Q}^{D+1} ~.~
\eea
Here, we can relabel the components of the $q$-exponent vector as $e_i = \Delta_b + s_i \in\mathbb{Q}$.

We note that the first component of the $(D+1)$-dimensional $q$-exponent vector is always the conformal weight $\Delta_b \in \mathbb{Q}$.
Having this in mind, we introduce a \textit{normalized $q$-exponent vector} for a $q$-series expanded to order $q^{\Delta_b+m_D}$ as follows, 
\beal{es30a1}
\mathbf{e}^{\prime} &=& \frac{1}{\Delta_b} \mathbf{e}
=
 \left[1, 1+\frac{s_1}{\Delta_b}, \dots, 1+\frac{s_D}{\Delta_b}\right] ~.~
\eea

We can also introduce a vector whose components measure the ``gap'' between consecutive exponents of the $q$-series.
We measure the gap as the ratio between consecutive exponents of the $q$-series, and define the following \textit{$q$-exponent gap vector}, 
\beal{es30a2}
\mathbf{r} &=& \left[\frac{e_1}{e_0}, \frac{e_2}{e_1}, \dots, \frac{e_D}{e_{D-1}}\right] 
\nn\\
&=&
\left[1+\frac{s_1}{\Delta_b}, \frac{\Delta_b+s_2}{\Delta_b+s_1}, \dots , \frac{\Delta_b+s_D}{\Delta_b+s_{D-1}}\right]
\in \mathbb{Q}^D
~,~
\nn\\
\eea
where each component is given by $r_i=\frac{e_i}{e_{i-1}}$. This data representation is designed to capture the rate of change of the $q$-series exponents (hence the name ``$r_i$''). 
We note here that there are other ways to represent the gaps in the $q$-series, which we plan to explore in future work.
We can normalize the $q$-exponent gap vector by dividing every vector component by the first component $\frac{e_1}{e_0}=1+\frac{s_1}{\Delta_b}$.
This ensures that all \textit{normalized $q$-exponent gap vectors} have $1$ as their first component,
\beal{es30a3}
\mathbf{r}^{\prime} = 
\frac{e_0}{e_1} 
\mathbf{r}
=
\left[
1, \frac{e_0 e_2}{e_1^2} ,\frac{e_0 e_3}{e_1 e_2}, \dots, \frac{e_0 e_D}{e_1 e_{D-1}}
\right]
\eea
and allows us to compare sparse spectra with dense spectra. Put differently, the normalization indicated with a prime normalized all BPS spectra to the same ``scale'' allowing us to compare many different QFT's focusing on the \textit{relative} features of BPS spectra. This will be useful in what follows, allowing us to explore which data representation shows a stronger PCA signal.

In this work, we propose the use of principal component analysis (PCA) to explore whether these vector representations of the $q$-series $\hat{Z}_b(q)$ capture any hidden structure of the BPS invariants.
In particular, we want to answer the following questions with the proposed use of PCA:
\begin{itemize}
\item Which vector representation encapsulates the most information about the $q$-series $\hat{Z}_b(q)$?
\item Which components of the vector representation statistically contribute most to the principal components?
\item Do the principal components tell us about relationships between different $q$-series $\hat{Z}_b(q)$ and the corresponding $3$-manifolds $Y$?
\end{itemize}
In order to answer these questions, let us first quickly review principal component analysis (PCA) in the context of vector representations for the $q$-series $\hat{Z}_b(q)$.
\\

\noindent
\textit{Principal Component Analysis (PCA).}
The aim of this work is to study the underlying structure of invariants encapsulated in the $q$-series of plumbed $3$-manifolds.
As discussed above, each $q$-series $\hat{Z}_b(q)$ can be associated with a vector representation $\mathbf{x}(\Gamma,\mathbf{m},b) \in \mathbb{R}^d$.
For the framings identified in \tref{t_tab01}, we have in total $N=20$ such $q$-series vectors $\mathbf{x}(\Gamma,\mathbf{m},b)$ living in a $d$-dimensional vector space. 
In order to study the hidden topological and geometric structures encapsulated by these vectors $\mathbf{x}(\Gamma,\mathbf{m},b)$, we make use of an unsupervised machine learning technique known as the \textit{principal component analysis (PCA)} \cite{pearson1901liii,jolliffe2002principal}.
Let us give here a brief overview of our proposed use of PCA for $\mathbf{x}(\Gamma,\mathbf{m},b) \in \mathbb{R}^d$.

Given the set of $N$ $q$-series vectors $\{\mathbf{x}_1, \dots, \mathbf{x}_N\}$, we can project the space $\mathbb{R}^d$ to a lower dimensional space $U=\mathbb{R}^k \subset \mathbb{R}^d$ with $k<d$.
The projected vectors are then of the form, 
\beal{es30a01}
\mathbf{z}_a = P_{\pi} \mathbf{x}_{a} = (z_{a1},\dots z_{ak}) \in U ~,~
\eea
where $a=1,\dots,N$. 
The projection matrix from $\mathbb{R}^d$ to $U$ takes the form $P_\pi= \mathbf{B} \mathbf{B}^\top$, where $\mathbf{B}=(\mathbf{b}_1,\dots,\mathbf{b}_k) \in \mathbb{R}^{d\times k}$ is the matrix of orthonormal basis vectors $\mathbf{b}_i$ for $U$ with $i=1,\dots,k$.

Under PCA, we want to find not just any subspace $U$ but an optimal subspace $\hat{U} \subset \mathbb{R}^d $ where statistically the maximum variance is achieved for the projected vectors $\mathbf{z}_a$. 
In other words, this means that we need to find the optimal subspace $\hat{U}$ with basis vectors $\mathbf{b}_1, \dots, \mathbf{b}_k$ such that when the original $q$-series vectors $\mathbf{x}_a \in \mathbb{R}^d$ are projected to $\hat{U}$, they have a maximized variance along $\mathbf{b}_1, \dots, \mathbf{b}_k$.
Here we note that the variance is a measure of the amount of vector component information that is preserved under the projection from $\mathbb{R}^d$ to $U$.

Let us assume that the $q$-series vector $\mathbf{x}_a \in \mathbb{R}^d$ has already been optimally projected along the first $r-1<k$ basis vectors $\mathbf{b}_1, \dots, \mathbf{b}_{r-1}$ of $\hat{U}$.
Given that the basis of $\hat{U}$ is orthonormal, the $r$-th coordinate $z_{ar}$ along $\mathbf{b}_r$ can be expressed as 
\beal{es30a02}
z_{ar} = \mathbf{b}_r^\top \mathbf{x}_a ~.~
\eea
We can calculate the variance of $z_{ar}$ along $\mathbf{b}_r$ over all $a$ as follows,
\beal{es30a05}
V(z_{r}) &=& \frac{1}{N} \sum_{a=1}^{N} z_{ar}^2 = \frac{1}{N} \sum_{a=1}^{N} (\mathbf{b}_r^\top \mathbf{x}_a)^2 
\nn\\
&=& \mathbf{b}_r^\top \mathbf{S} \mathbf{b}_r ~,~
\eea
where $\mathbf{S}$ is the $q$-series covariance matrix over the original set of $q$-series vectors $\{\mathbf{x}_1, \dots, \mathbf{x}_N\}$ in $\mathbb{R}^d$.
The $q$-series covariance matrix in $d$-dimensions is defined as
\beal{es30a06}
\mathbf{S} = \frac{1}{N} \sum_{a=1}^{N} \mathbf{x}_a \mathbf{x}_a^\top = \frac{1}{N} \mathbf{X} \mathbf{X}^\top
~,~
\eea
where $\mathbf{X} = (\mathbf{x}_1, \dots, \mathbf{x}_N)$ is a $d\times N$ matrix and $\mathbf{S}$ is a $d\times d$ matrix.

Using the $q$-series covariance matrix, the optimization problem of finding $\hat{U}$ along $\mathbf{b}_r$ can be summarized by the Lagrangian function 
\beal{es30a07}
\mathcal{L}(\mathbf{b}_r,\lambda) = \mathbf{b}_r^\top \mathbf{S} \mathbf{b}_r + \lambda_r (1- \mathbf{b}_r^\top \mathbf{b}_r) ~,~
\eea
where $\lambda$ is the Lagrange multiplier for the constraint that the basis vectors of $\hat{U}$ are orthonormal, $\mathbf{b}_r^\top \mathbf{b}_r = 1$.
Solving this constrained optimization problem leads to the following eigenvalue equation 
\beal{es30a10}
\mathbf{S} \mathbf{b}_r = \lambda_r \mathbf{b}_r ~,~
\eea
where $\lambda_r$ is now the eigenvalue of the $q$-series covariance matrix $\mathbf{S}$ with the corresponding eigenvector $\mathbf{b}_r$.
The variance $V(z_{r})$ of vector components along $\mathbf{b}_r$ in $\hat{U}$ is then given by the eigenvalue $\lambda_r$.

Overall, this implies that in order to obtain the optimal $k$-dimensional feature subspace $\hat{U}$ from the collection of $N$ $q$-series vectors $\{\mathbf{x}_1, \dots, \mathbf{x}_N\}$ in $\mathbb{R}^d$, we have to calculate the $k$ largest eigenvalues $\lambda_i$ and the corresponding eigenvectors $\mathbf{b}_i$ of the $q$-series covariance matrix $\mathbf{S}$ in order to obtain $\hat{U}$. 
We note here that if $N < d$, meaning the number of $q$-series vectors is smaller than the vector dimension itself, then the $q$-series covariance matrix $\mathbf{S}$ will not be full rank. 
This means the number of non-zero eigenvalues $\lambda_i$ is at most $N-1$.

\begin{figure*}[ht!!]
\begin{center}
\resizebox{0.98\hsize}{!}{
\includegraphics[height=5cm]{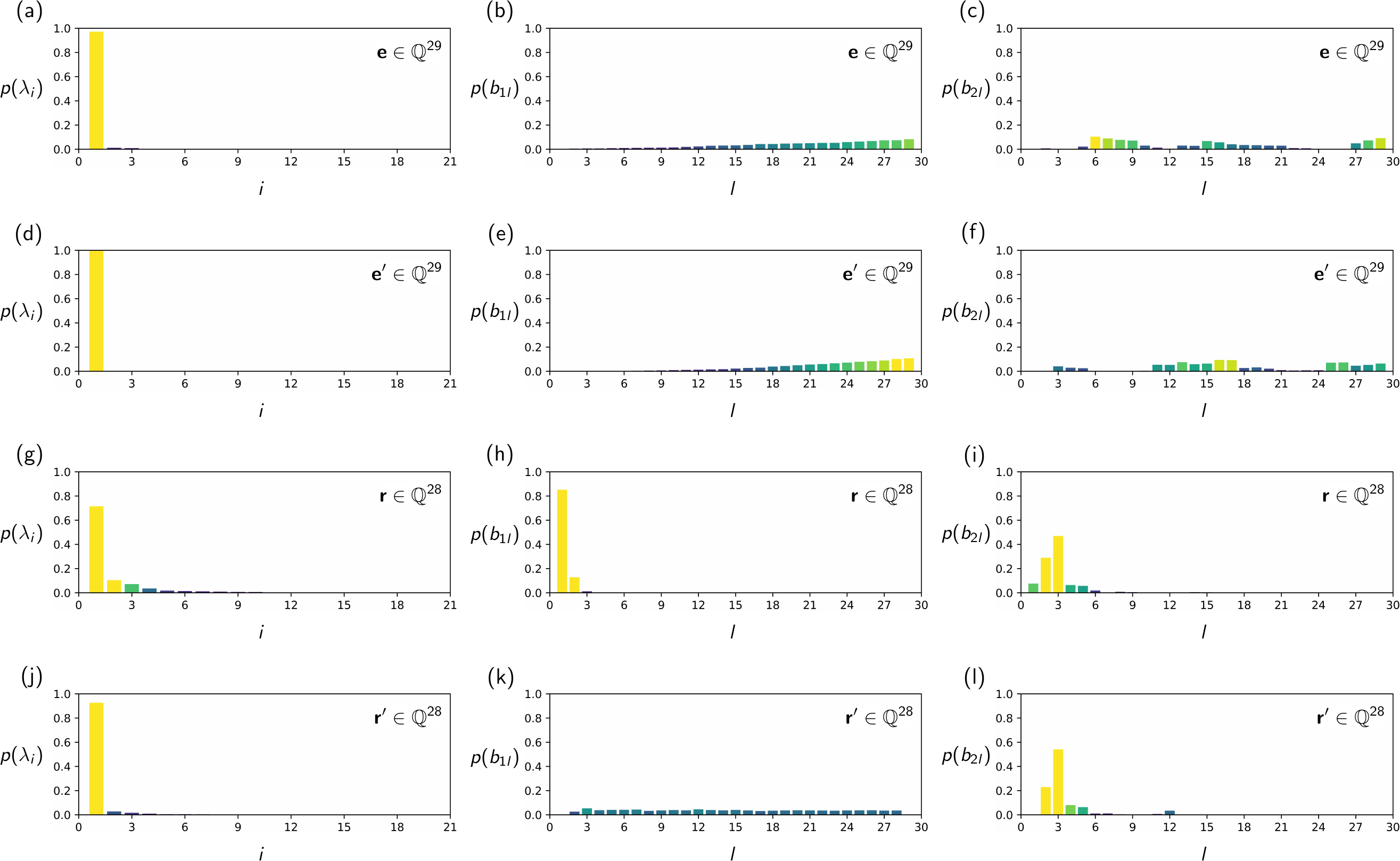} 
}
\caption{
Summary of the exponent relevance analysis using PCA. 
We note that the first principal component $\mathbf{b}_1$ covers the vast majority of the proportional variance of the $q$-series vectors as shown in (a), (d), (g) and (j).
This implies that for all vector representations, the vector spaces dimensionally reduce effectively to a 1-dimensional subspace.
A closer look reveals that the components of $q$-exponent vector $\mathbf{e}$ and the corresponding normalized vector $\mathbf{e}^\prime$ do contribute to the first principal component when the exponent values becomes larger, as expected and shown in (b) and (e).
For the $q$-exponent gap vector $\mathbf{r}$, we see however that only the first 2 components significantly contribute to the first principal exponent as shown in (h). When normalized, this contribution is evenly distributed to all the components of the normalized $q$-exponent gap vector $\mathbf{r}^\prime$ as shown in (k).
Similar observations can be made in (c), (f), (i) and (l) for the second principal components, whose contribution is significantly less.
\label{f_res01}}
 \end{center}
 \end{figure*}

In \cite{Seong:2023njx}, the $2$-dimensional subspace $\hat{U}$ obtained from coamoeba vectors $\mathbf{x}_a$ associated to choices of complex structure moduli in the Calabi-Yau mirror description of the cone over the zeroth Hirzebruch surface $F_0$ was interpreted as a \textit{phase space} of the corresponding $4$-dimensional supersymmetric gauge theories related by Seiberg duality. 
In our study here, we interpret $\hat{U}$ obtained by PCA as a \textit{latent space} of 
the set of all possible $q$-series and the associated $3$-manifolds constructed from a specific plumbing graph $\Gamma$ with a collection of associated framing numbers.
We argue in this work that this latent space 
exhibits interesting relationships between the $3$-manifolds corresponding to the `H-shaped' plumbing graph in \fref{f_fig01} and Spin$^c$ structure $b=(0,0,0,0,0,0)$, reminiscent of a QFT phase diagram. 

Beyond the latent space itself, we make observations regarding the eigenvectors $\mathbf{b}_i$ corresponding to the largest eigenvalues $\lambda_i$ in the $q$-series covariance matrix $\mathbf{S}$ and the $q$-series $Z_b(q)$ of the $3$-manifold itself. 
Let us summarize the role played by the eigenvectors $\mathbf{b}_i$ in relation to the input vectors $\{\mathbf{x}_1,\dots,\mathbf{x}_N\}$ in the PCA.
\\

\noindent
\textit{Meaning of the Eigenvectors and Eigenvalues of the $q$-Series Covariance Matrix.}
The eigenvectors $\mathbf{b}_i$ of the $q$-series covariance matrix $\mathbf{S}$ have $d$ components of the form $b_{il}$, where $i=1,\dots, r$ labels the principal components and $l=1,\dots, d$ labels the eigenvector components.
Since we assume that the eigenvectors $\mathbf{b}_i$ are orthonormal, the components have to satisfy,
\beal{es40a01}
\sum_{l=1}^{d} (b_{il})^2 = 1 ~,~
\eea
Given that the squares of the components sum up to 1, we can interpret them as relative percentage measures
\beal{es40a02}
p(b_{il}) = (b_{il})^2 \times 100\% ~.~
\eea
This measures the \textit{proportional component contribution} of the $l$-th component $b_{il}$ of the $q$-series vector towards the $i$-th principal component $\mathbf{b}_i$. 
The higher the value for $p(b_{il})$, the more relevant becomes the corresponding component in the $q$-series vector for the principal component analysis, storing more information relative to other components of the $q$-series vector about the $q$-series under dimensional reduction.

In addition to the percentage component contribution in \eref{es40a02}, 
we can also define the \textit{proportional variance} related to the $i$-th principal component as follows, 
\beal{es40a05}
p(\lambda_i) = \frac{\lambda_i}{\sum_{j=1}^{d} \lambda_j} \times 100\% ~.~
\eea
Here, the proportional variance $p(\lambda_i)$ measures the proportional importance of the corresponding principal component and the amount of information captured by the principal component from the set of $q$-series vectors $\mathbf{x}_1, \dots, \mathbf{x}_N$.

In the following section, we make use of principal component analysis in order to measure the significance of certain terms in the $q$-series expansion, when they are represented as vectors.
By obtaining the proportional variance $p(\lambda_i)$, we can accurately measure whether the $q$-series for the same plumbing graph but different framings exhibit any similarities. 
These similarities are parameterized by principal components $\mathbf{b}_i$, where the proportional component contribution $p(b_{il})$ tells us how much each exponent in the $q$-series contributes to a certain principal component $\mathbf{b}_i$. 
\\

\section{Results}

\begin{figure}[ht!!]
\begin{center}
\resizebox{0.9\hsize}{!}{
\includegraphics[height=5cm]{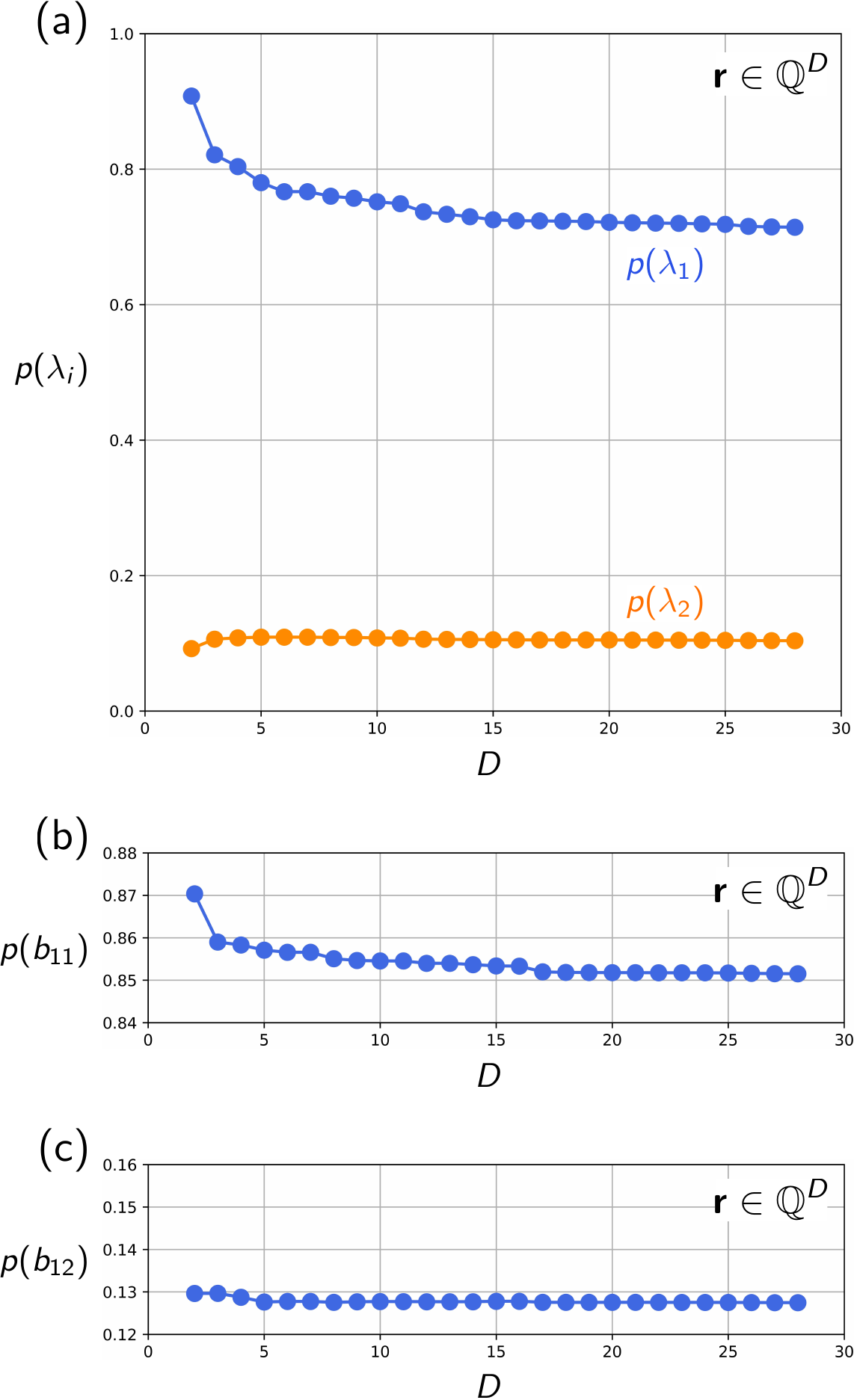} 
}
\caption{
The change in proportional variance $p(\lambda_i)$ for the first 2 principal components $\mathbf{b}_1$ and $\mathbf{b}_2$ when one takes only the first $D$ components $r_1, \dots , r_D$ of the $q$-exponent gap vectors $\mathbf{r}$ is shown in (a).
Focusing on the first principal component $\mathbf{b}_1$, the relative contributions to $\mathbf{b}_1$ of the vector components $r_1$ and $r_2$ under changing the number of components of $\mathbf{r}$ are shown in (b) and (c), respectively.
\label{f_res02}}
 \end{center}
 \end{figure}

For the `H-shaped' plumbing graph in \fref{f_fig01}, 
we make use of the classification of inequivalent framing numbers in \tref{t_tab01} in order to calculate the corresponding $q$-series $\hat{Z}_b(q)$ with Spin$^c$ structure $b=[0,0,0,0,0,0]$.
\tref{t_tab02} and \tref{t_tab03} show the $q$-series expansions with their corresponding framing numbers $\mathbf{m}$.
Up to a very large permutation symmetry of graph labels, we have $N=20$ distinct $q$-series.
We expand the $q$-series such that in total the corresponding $q$-exponent vectors $\mathbf{e}$ from \eref{es29a01b2} are all $(D+1)$-dimensional, with $D=28$. 
As a result, the corresponding $q$-exponent gap vectors $\mathbf{r}$ defined in \eref{es30a2} are all $28$-dimensional.

Given the vector representations of the $q$-series in terms of the $q$-exponent vectors $\mathbf{e}$, normalized $q$-exponent vectors $\mathbf{e}^\prime$, the $q$-exponent gap vector $\mathbf{r}$ and the normalized $q$-exponent gap vector $\mathbf{r}^\prime$, 
the resulting principal component analysis of these 4 vector spaces gives several interesting results.

First of all, for any vector representation, the first principal component covers the vast majority of the variance of the datasets.
For the 4 vector spaces, we obtain
\beal{es50a1} 
&
p(\lambda_1; \mathbf{e})
= 
97.111 \%
~,~
p(\lambda_1; \mathbf{e}^\prime)
= 99.815 \%
~,~
&
\nn\\
&
p(\lambda_1; \mathbf{r}) = 71.426 \%
~,~
p(\lambda_1; \mathbf{r}^\prime) = 92.584 \%
~,~
&
\nn\\
\eea
as illustrated in \fref{f_res01}(a), (d), (g) and (j), respectively.
This shows that the vector spaces formed by the vector representations of the $q$-series are effectively all $1$-dimensional. 
Only for the $q$-exponent gap vectors $\mathbf{r}$, we see that more principal components cover the variance of the dataset, 
\beal{es50a2}
p(\lambda_2; \mathbf{r}) =
10.385 \%
~,~
p(\lambda_3; \mathbf{r}) =
7.132 \%
~,~
\eea
as illustrated in \fref{f_res01} (g). 
In total, we have $N-1=19$ non-trivial principal components $\mathbf{b}_i$ with non-zero eigenvalues $\lambda_i$ since $N<D+1$ for the exponent vectors and $N<D$ for the rate of exponent change vectors. 

Without loss of generality, let us focus on the first principal component $\mathbf{b}_1$ for all 4 sets of vector representations.
By looking at the components $b_{il}$ of the principal components, we can calculate the proportional component contribution of the $l$-th component of the vector for the $i$-th principal component $\mathbf{b}_i$, as summarized in \eref{es40a02}.

We note that for $q$-exponent vectors $\mathbf{e}$ and $\mathbf{e}^\prime$, the components of the vectors contribute more with higher component index $l$.
This is because for both the original and normalized $q$-exponent vectors, the components exhibit larger variance for increasing values of the $q$-series exponents. As a result, larger component values in $\mathbf{e}$ and $\mathbf{e}^\prime$ lead to increased contribution towards the dominating first principal component as shown in \fref{f_res01}(b) and (e).

In comparison, when we consider the $q$-exponent gap vector $\mathbf{r}$, we make an interesting observation that the first couple of components contribute significantly more towards $\mathbf{b}_1$, than all the other remaining components.
The first 4 components have the following proportional component contributions,
\beal{es50a10}
&
p(b_{11}; \mathbf{r}) = 
85.153 \%
~,~
p(b_{12}; \mathbf{r}) = 
12.747 \%
~,~
&
\nn\\
&
p(b_{13}; \mathbf{r}) = 
1.125 \%
~,~
p(b_{14}; \mathbf{r}) = 
0.181 \%
~,~
&
\eea
which shows that over $97\%$ of the contribution towards the first principal component $\mathbf{b}_1$ comes from the first two components of $\mathbf{r}$ given by 
\beal{es50a12}
r_1 = \frac{e_1}{e_0} = \frac{\Delta_b + s_1}{\Delta_b} ~,~
r_2 = \frac{e_2}{e_1} = \frac{\Delta_b + s_2}{\Delta_b + s_1} ~.~
\eea
This is illustrated in \fref{f_res01} (h).
We note here that this is a result expected for close to evenly distributed ordered sequences, which appears to be the case for the exponents of the collection of $q$-series studied in our work.
A characteristic feature here for the $q$-series investigated in our work is the observation that the first $2$ gaps measured by ratios $r_1$ and $r_2$ are significantly more relevant than any other ratios that follow.  
We hope to investigate these properties further with a larger collection of $q$-series in future work.

When we consider the normalized $q$-exponent gap vector $\mathbf{r}^\prime$, every component in $\mathbf{r}$ is divided by its first component.
Given that the first component $r_1$ in $\mathbf{r}$ is also the most significantly contributing component with $p(b_{11}; \mathbf{r}) = 85.153 \%$, 
dividing every component by it evenly distributes the contribution on every component of the normalized vector $\mathbf{r}^\prime$.
As a result, the components $\mathbf{r}^\prime$ have all proportional component contributions in the range
\beal{es50a15}
2\% <p(b_{1j}; \mathbf{r} ^\prime) < 6\% ~,~
\eea
as illustrated in \fref{f_res01}(k).

We can measure how the proportional component contributions of $r_1$ and $r_2$ towards the first principal component $\mathbf{b}_1$ change when one varies the overall number of components of $\mathbf{r}$.
When one varies the overall number of components of $\mathbf{r}$, we effectively vary the number of terms in the $q$-series expansion that contribute towards the principal component analysis. 
From the results in \fref{f_res02}(b) and (c), we see that $p(b_{11}; \mathbf{r})$ and $p(b_{12}; \mathbf{r})$ converge towards a lower bound when one increases the dimensionality of the vectors $\mathbf{r}$. 
This implies that including more terms in the $q$-series expansion does not change our observation that the first two components in \eref{es50a12} of the $q$-exponent gap vector $\mathbf{r}$ contribute vastly more than any other components of $\mathbf{r}$, even if more terms in the $q$-series expansion are included in the analysis. 

We can also track how the proportional variance $p(\lambda_1)$ of the first principal component $\mathbf{b}_1$ varies when one includes more components to the vector $\mathbf{r}$. 
As shown in \fref{f_res02}(a), we see that the proportional variance $p(\lambda_1)$ reaches asymptotically a lower bound above $70 \%$, implying that the space of $q$-series vector representations is effectively 1-dimensional independent on how many terms in the $q$-series expansion are included in the principal component analysis.
\\

\section{Discussions}

In this work, we have studied the $q$-series $\hat{Z}_b(q; Y)$ 
for a family of $3$-manifolds $Y$ associated to the same H-shaped plumbing graph in \fref{f_fig01} with different framings $\mathbf{m}$ as summarized in \tref{t_tab01}.
By identifying the gap between $q$-series exponents as the ratio between two consecutive exponents in the $q$-series, we have represented the $q$-series for particular $3$-manifolds $Y$ as finite-dimensional vectors whose components are made of the first few gaps of the corresponding $q$-series. 
By employing machine learning methods for dimensional reduction of the input data --- specifically, the principal component analysis (PCA) --- we are able to compute the statistical significance of each gap in the spectrum of BPS states captured by the $q$-series.
The statistical significance is given by the saliency measure of the features of the input data, which is computable for any explainable machine learning model. In our case, these measures are given by the components of the eigenvectors of the covariance matrix obtained from PCA which correspond to gaps in the collection of input $q$-series vectors.

Our work shows that for the family of $q$-series $\hat{Z}_b(q; Y)$ 
corresponding to $3$-manifolds $Y$ associated to the same H-shaped plumbing graph with framing numbers given in \tref{t_tab01}, 
the first two gaps in the $q$-series are statistically more significant than all other gaps that follow them in the $q$-series. 
This indicates that BPS degeneracies of the corresponding $3d$ $\mathcal{N}=2$ supersymmetric theories are associated to statistically more significant exponent gaps in the corresponding $q$-series than others. 
We expect this somewhat surprising result to be more general and hold true for other families of $q$-series associated with plumbing graphs of other shapes.

Our work also illustrates the potential of explainable machine learning techniques that enable the measure of saliency scores for features in input datasets.
These measures in turn allow us, as illustrated in this work, to discover new structures in the spectra of states in quantum systems and
in the counting of enumerative invariants characterizing geometries and topologies associated with these quantum systems.
As part of the wider study of quantum systems and their spectra of quantum states,
our work illustrates that for BPS spectra of $3d$ $\mathcal{N}=2$ supersymmetric theories, 
BPS states counted by the first few terms of the $q$-series are statistically more significant compared to the later terms in the $q$-series.
We expect to report more findings in this direction in the near future. 
\\

\acknowledgments

The authors would like to thank Miranda Cheng, Hee-Joong Chung, Shimal Harichurn, Arnav S. Kabra, Davide Passaro, Fabian Ruehle, and Josef Svoboda for discussions and comments.
The work of S.G. is supported in part by a Simons Collaboration Grant on New Structures in Low-Dimensional Topology, by the NSF grant DMS-2245099, and by the U.S. Department of Energy, Office of Science, Office of High Energy Physics, under Award No. DE-SC0011632.
R.-K. S.  is supported by a Basic Research Grant of the National Research Foundation of Korea (NRF-2022R1F1A1073128).
He is also supported by a Start-up Research Grant for new faculty at UNIST (1.210139.01) and a UNIST AI Incubator Grant (1.240022.01).  
He is also partly supported by the BK21 Program (``Next Generation Education Program for Mathematical Sciences'', 4299990414089) funded by the Ministry of Education in Korea and the National Research Foundation of Korea (NRF).
\\


\bibliographystyle{jhep}
\bibliography{mybib}

\end{document}